\documentclass[twocolumn]{aastex701}

\DeclareRobustCommand{\ion}[2]{\textup{#1\,\textsc{\lowercase{#2}}}}
\usepackage{amsmath}
\definecolor{jmdsscolor}{RGB}{200,0,0}

\begin{document}

\title{Chromospheric magnetic field extrapolations reveal the flux-rope configuration of a solar filament}

\author[0000-0002-9309-2981]{Robert Jarolim}
\affiliation{High Altitude Observatory, NSF National Center for Atmospheric Research, USA}
\affiliation{Institute of Physics, University of Graz, Austria}
\email{rjarolim@ucar.edu}

\author[0000-0002-3009-295X]{Jo{\~a}o M. da Silva Santos}
\affiliation{National Solar Observatory, 3665 Discovery Drive, Boulder, CO 80303, USA}
\email{}

\author[0000-0001-5850-3119]{Matthias Rempel}
\affiliation{High Altitude Observatory, NSF National Center for Atmospheric Research, Boulder, CO, USA}
\email{}

\author[0000-0002-0049-4798]{Marianna B. Kors\'os}
\affil{School of Electrical and Electronic Engineering, University of Sheffield, Amy Johnson Building, Portabello Street, Sheffield, S1 3JD, UK}
\affil{Department of Astronomy, E\"otv\"os Lor\'and University,  P\'azm\'any P\'eter s\'et\'any 1/A, H-1112 Budapest, Hungary}
\affil{Gyula Bay Zolt\'an Solar Observatory (GSO), Hungarian Solar Physics Foundation (HSPF), Pet\H{o}fi t\'er 3, H-5700 Gyula, Hungary}
\email{}

\author[0000-0003-3439-4127]{Robertus Erd{\'e}lyi}
\affiliation{Solar Physics and Space Plasma Research Centre (SP2RC), School of Mathematics and Statistics, The University of Sheffield, Sheffield, UK}
\affiliation{Department of Astronomy, E{\"o}tv{\"o}s Lor{\'a}nd University, Budapest, Hungary}
\email{}

\author[0000-0003-2073-002X]{Astrid Veronig}
\affiliation{Institute of Physics, University of Graz, Graz, Austria}
\affiliation{Kanzelh{\"o}he Observatory for Solar and Environmental Research, University of Graz, Graz, Austria}
\email{}

\author[0000-0002-3606-161X]{Szabolcs So{\'o}s}
\affiliation{Department of Astronomy, E{\"o}tv{\"o}s Lor{\'a}nd University, Budapest, Hungary}
\email{}

\author[0000-0003-2760-2311]{David Kuridze}
\affiliation{National Solar Observatory, 3665 Discovery Drive, Boulder, CO 80303, USA}
\affiliation{Georgian National Astrophysical Observatory, Abastumani, 0301, Georgia}
\email{}

\begin{abstract}

Solar eruptions are powered by the release of magnetic energy stored in the lower solar atmosphere, but the pre-eruptive magnetic configuration of filament channels remains difficult to determine. A central question is whether this energy is stored in a pre-existing magnetic flux rope or in a sheared arcade that forms a flux rope only during eruption. Resolving this ambiguity is critical for identifying instability thresholds and eruption triggers, yet photosphere-based extrapolations often provide insufficient constraints on the three-dimensional coronal field.
Here, we introduce a data-driven magnetic field extrapolation framework that combines photospheric and chromospheric vector magnetograms in a unified multi-height optimization, while accounting for variable chromospheric formation heights and the 180° azimuthal ambiguity. Tests with radiative magnetohydrodynamic simulations show that photosphere-only extrapolations can misidentify the pre-eruptive magnetic configuration, whereas chromospheric vector constraints recover the three-dimensional structure substantially more accurately.
Applied to multi-line spectropolarimetric observations of an active region filament obtained with the Swedish Solar Telescope, the method reveals a reconstructed magnetic field consistent with a pre-eruptive flux-rope configuration. These results show that chromospheric vector magnetic measurements can provide decisive constraints on filament magnetic configuration and open a path toward diagnosing magnetic-energy storage and instability in eruptive solar active regions.
\end{abstract}


\section{Introduction}

The structure and connectivity of the solar magnetic field govern solar eruptions, during which large amounts of magnetic energy are released on timescales of seconds to minutes. By contrast, the buildup of magnetic energy occurs over hours to days, making the pre-eruptive magnetic configuration critical for understanding how energy is stored in the solar atmosphere and evolves toward eruption \citep{2018SSRv..214...46G}. In spatially confined regions, such as active regions and quiet-Sun filaments, highly twisted magnetic fields can accumulate substantial magnetic energy and become prone to instability. Although modeling and observations provide plausible explanations for the supporting magnetic configuration, sparse measurements and significant uncertainties still leave the true magnetic configuration poorly constrained.

Magnetohydrodynamic (MHD) theory and simulations suggest two competing magnetic configurations that can support suspended plasma in the solar atmosphere: a \textbf{sheared magnetic arcade (SMA)} and a \textbf{magnetic flux rope (MFR)} \citep{2018LRSP...15....7G}. In an MFR, field lines wind around a common central axis, whereas in an SMA they connect opposite polarity regions without forming a coherent twisted core. Three-dimensional modeling has shown that shearing motions along the polarity inversion line can produce prominence-supporting magnetic dips in arcade-like fields without requiring a pre-existing flux rope \citep{Antiochos1994,Aulanier2002}. Conversely, flux ropes may form through magnetic reconnection within an SMA, often associated with flux cancellation at the photosphere, or through the emergence of already twisted magnetic flux, which can appear as rotational or twisting motions in photospheric observations \citep{1989ApJ...343..971V,2000ApJ...539..954D,1999ApJ...518L..57A,2004ApJ...609.1123F}. Eruption and thermodynamic MHD simulations connect these scenarios by showing that tether-cutting and bald-patch reconnection can transform an SMA into an erupting MFR, while prominence condensations can form in the dipped fields of pre-existing flux ropes \citep{2001ApJ...552..833M,2010ApJ...708..314A,2012ApJ...758...60F,2017ApJ...844...26F,2023ApJ...950L...3C,rempel2023ApJmfr_sma}.

Observations provide strong but indirect evidence for magnetic flux ropes in solar eruptions. White-light CME morphologies are frequently consistent with erupting flux ropes \citep{2013SoPh..284..179V}, while pre-eruptive candidates are inferred from prominence cavities, sigmoids, and hot EUV channels \citep{2010ApJ...724.1133G, 2014LRSP...11....1P}. However, hot-channel and on-disk observations suggest that the erupting MFR may either pre-exist or form dynamically during the eruption \citep{2020A&A...642A.109N}. Since these diagnostics are affected by projection, optically thin emission, thermal response effects, and the fact that both SMAs and MFRs can support prominence material, imaging alone cannot uniquely determine the magnetic configuration. Distinguishing between an SMA and an MFR therefore requires constraints on the three-dimensional magnetic field structure. In particular, \citet{2020SSRv..216..131P} emphasized that combining multi-viewpoint, multi-thermal coronal observations with multi-height vector magnetic field measurements is key to distinguishing between pre-eruptive magnetic configurations.

The solar magnetic field is routinely measured in photospheric layers. By contrast, measurements of the coronal magnetic field remain challenging because of thermal line broadening and the optically thin medium. Although the first coronal magnetic field observations have now been obtained \citep{2007Sci...317.1192T, 2019ApJ...874..126K, 2020Sci...369..694Y, 2024SciA...10.1604S, 2024Sci...386...76Y}, they provide only two-dimensional, line-of-sight-integrated information and are therefore insufficient for determining the three-dimensional magnetic configuration. Likewise, photospheric observations alone sample only a single lower-atmosphere boundary and do not reliably distinguish between different three-dimensional magnetic configurations.

Magnetic field extrapolations and data-driven MHD simulations provide the main route for inferring the 3D magnetic field of the solar atmosphere from routinely available vector magnetograms. Nonlinear force-free (NLFF) extrapolations can recover magnetic connectivity, quasi-separatrix layers (QSLs), and pre-eruptive flux-rope-like structures \citep{wiegelmann2021sffmf,Korsos2024preeruption,2016RAA....16...15J, 2020ApJ...896..119K}, with synthetic tests demonstrating their ability to recover flux-rope-like configurations such as Titov--Démoulin magnetic flux ropes \citep{1999A&A...351..707T,2016RAA....16...15J}. Time-dependent data-constrained and data-driven MHD simulations extend this inference to dynamic eruption scenarios \citep{jiang2022data,2024RvMPP...8...29G,2024RvMPP...8...27S,2025Sci...388.1306D, 2020ApJ...903..129P}. However, inference of magnetic connectivity remains sensitive to boundary conditions and modeling assumptions. Photospheric magnetograms provide routine vector-field measurements, but the photosphere is generally not force-free. Chromospheric measurements are more compatible with the force-free approximation \citep[e.g.,][]{1995ApJ...439..474M,yelles2012chromospheric_extrapolation,wiegelmann2021sffmf, 2022JSWSC..12....2E}, but sample corrugated formation-height surfaces due to spatial variations in line formation height \citep[e.g.,][]{2019A&A...631A..33B,2020A&A...642A.210M}. In addition, spectropolarimetric vector-field inference is inherently subject to the 180$^\circ$ ambiguity of the transverse-field azimuth. These challenges motivate extrapolation methods that combine photospheric and chromospheric constraints while accounting for uncertain geometric heights and ambiguous transverse-field directions.

In this study, we address the ambiguity in distinguishing magnetic flux ropes from strongly sheared arcades using multi-height magnetic field extrapolations based on both photospheric and chromospheric magnetic field measurements \citep{jarolim2024multi_height}. We provide three primary contributions: (1) we validate the reliability of NLFF extrapolations in distinguishing between sheared magnetic arcade and magnetic flux rope configurations, (2) we identify the magnetic configuration of a solar filament and demonstrate the relevance of chromospheric magnetic field measurements, and (3) we extend the extrapolation framework to intrinsically account for azimuthal ambiguity, improve convergence times, and enhance extrapolation quality. We demonstrate the benefit of a coupled inference framework in which magnetic field extrapolation, geometrical-height estimation of corrugated surfaces, and intrinsic disambiguation jointly provide more robust constraints on the three-dimensional magnetic structure.

\section{Methods} 
\label{sec:method}

We build on the physics-informed neural network (PINN)-based magnetic field extrapolation framework introduced by \citet{jarolim2023nf2} (Neural Network Force-Free extrapolations; NF2), including its extensions to multi-height extrapolations \citep{jarolim2024multi_height} and the vector-potential formulation \citep{Jarolim2024ar13664}.
In this framework, a neural network represents the magnetic field as a continuous function of position. Given spatial coordinates $(x,y,z)$, the network returns the magnetic field vector $\mathbf{B}(x,y,z)$, or, in the vector-potential formulation, a vector potential $\mathbf{A}(x,y,z)$ from which the magnetic field is derived as
\begin{equation}
    \mathbf{B} = \nabla \times \mathbf{A}.
\end{equation}
This construction satisfies $\nabla \cdot \mathbf{B}=0$ by definition. The remaining force-free condition,
\begin{equation}
    \left(\nabla \times \mathbf{B}\right) \times \mathbf{B} = 0,
\end{equation}
is enforced through the optimization objective, together with losses that constrain the solution to the observed magnetic field at the lower boundary and to a potential field approximation at the side and top boundaries.

Recent neural-operator approaches have demonstrated real-time or near-real-time NLFF extrapolations by learning mappings from photospheric vector magnetograms to coronal magnetic fields \citep{2025ApJS..277...54J,2025ApJ...993..146Y,2024SpWea..2203875Z}. These models are complementary to PINN-based extrapolations: neural operators offer rapid inference once trained, but depend on representative training data, fixed data conventions, and the validity of the learned distribution. In contrast, the PINN formulation used here solves each extrapolation as an independent inverse problem, without requiring pre-training on reference extrapolations. This makes the method flexible with respect to domain size, instrumental setup, and available magnetic constraints, enabling applications to chromospheric magnetic field observations \citep{jarolim2024multi_height}, heterogeneous datasets \citep{2025ApJ...985..157D}, and large domains that are difficult to treat with conventional solvers \citep{Purkhart2025filament}.

Here, we extend the NF2 framework to fully exploit photospheric and chromospheric vector magnetic field measurements. The main methodological extensions are:
\begin{enumerate}
    \item intrinsic treatment of the 180$^\circ$ azimuthal ambiguity in vector magnetograms,
    \item improved optimization and convergence through height scaling, and
    \item positional encoding to enhance the representation of small-scale magnetic structure.
\end{enumerate}
Together, these extensions couple magnetic field extrapolation, chromospheric height inference, and azimuthal disambiguation within a single optimization framework. This provides a consistent basis for testing whether NLFF extrapolations can distinguish sheared magnetic arcades from magnetic flux ropes.

\subsection{Training}
\label{sec:training}

To enable automatic disambiguation, we reformulate the loss function such that it is invariant under a $180^{\circ}$ change in the azimuth orientation. We split the boundary loss into horizontal ($x,y$) and vertical ($z$) components. For the horizontal magnetic field, we optimize
\begin{equation}
    \mathcal{L}_{\mathbf{B}_{xy}} 
= \|\mathbf{B}_{xy}^{\text{true}}\|^2 
+ \|\mathbf{B}_{xy}\|^2 
- 2 \left| \mathbf{B}_{xy} \cdot \mathbf{B}_{xy}^{\text{true}} \right| \,,
\end{equation}
where $\mathbf{B}_{xy}^{\text{true}}$ denotes the observed horizontal magnetic field. For the vertical component, we directly match the reference value,
\begin{equation}
    \mathcal{L}_{B_z} = \left( B_z - B_z^{\text{true}} \right)^2 \,,
\end{equation}
where $B_z^{\text{true}}$ denotes the observed line-of-sight field component. We then combine both terms and assign a larger weight to the vertical component,
\begin{equation}
    \label{eq:bottom}
    \mathcal{L}_{\rm bottom} = 2 \left( w_{\text{los}} \, \mathcal{L}_{B_z} + (1 - w_{\text{los}})\, \mathcal{L}_{\mathbf{B}_{xy}} \right) \,,
\end{equation}
where we use $w_{\text{los}}=0.7$ for all extrapolations. The prefactor 2 keeps the overall scale comparable to the standard boundary loss. In particular, for $w_{\text{los}}=0.5$ and an unambiguous horizontal field, this expression reduces to the usual squared-error form, $\mathcal{L}_{\rm bottom}=\mathcal{L}_{B_z}+\mathcal{L}_{\mathbf{B}_{xy}}=\|\mathbf{B}-\mathbf{B}^{\text{true}}\|^2$.
For the side and top boundaries, we match the full vector components to the potential field solution 
\begin{equation}
    \label{eq:boundary}
    \mathcal{L}_{\rm potential-boundary}=\|\mathbf{B}-\mathbf{B}_{\text{potential}}\|^2 \,.
\end{equation}

Photospheric input data are approximated as flat and serve as the lower boundary condition, while chromospheric observations are intrinsically remapped using a second neural network that translates the initially guessed height $z'$ to the model height $z$ \citep[cf.][]{jarolim2024multi_height}. This remapping is integrated as part of the optimization.
We allow for variation between 0 and 20 Mm. In addition, we add a soft regularization toward the initial height guess,
\begin{equation}
    \label{eq:height-regularization}
    \mathcal{L}_{\rm height} = \frac{\left(z - z'\right)^2}{z'^2 + \epsilon} \, ,
\end{equation}
where $\epsilon=10^{-7}$ prevents division by values close to zero.

With the use of the vector potential $\mathbf{A}$, the divergence-free condition $\nabla \cdot \mathbf{B} = 0$ is satisfied up to numerical precision, and the primary physics constraint reduces to the force-free condition,
\begin{equation}
    \label{eq:force-free}
    \mathcal{L}_{\rm force-free} = \frac{\left\| (\nabla \times \mathbf{B}) \times \mathbf{B} \right\|^2}{\left\|\mathbf{B} \right\|^2 + \epsilon} \,.
\end{equation}
This term penalizes residual Lorentz forces throughout the extrapolation volume. Here, $\epsilon=10^{-7}$ is a small positive constant that prevents numerical instabilities in weak-field regions and avoids division by values close to zero.

We improve convergence stability and the overall quality of the extrapolations in two ways: (1) by suppressing spurious strong currents during the initial optimization phase and (2) by applying a height-dependent scaling to the physics losses.

First, as can be seen from Eq. \ref{eq:force-free}, the force-free loss can in principle be reduced by increasing $\left\|\mathbf{B} \right\|^2$ while decreasing $\left\| (\nabla \times \mathbf{B}) \times \mathbf{B} \right\|^2$. Although this behavior is constrained by the boundary terms in Eqs. \ref{eq:bottom} and \ref{eq:boundary}, the early optimization stage can still produce isolated current systems that are only weakly coupled to the imposed boundary conditions. To suppress these transients, we add an auxiliary current-minimization term,
\begin{equation}
    \label{eq:current-init}
    \mathcal{L}_{\rm current} = \left\| \nabla \times \mathbf{B} \right\|^2 \,.
\end{equation}
This term is included only during the first 5,000 optimization steps.

Second, because the magnetic field strength generally decreases with height, the optimization would otherwise be biased toward the lower layers and leave the upper part of the extrapolation volume less constrained. In each update step, we therefore sample 512 random $z$ layers and 128 random $(x, y)$ coordinate points per layer, corresponding to a total batch size of 65,536 volume samples. From these samples we estimate a height-dependent normalization,
\begin{equation}
    \label{eq:height-scale}
    w(z_i) = \left( \frac{1}{N_{xy}} \sum_{j=1}^{N_{xy}} \left\| \mathbf{B}(x_j, y_j, z_i) \right\|^2 + \epsilon \right)^{-1} \,.
\end{equation}
This normalization is applied to both $\mathcal{L}_{\rm force-free}$ and $\mathcal{L}_{\rm current}$. This scaling maintains a more uniform optimization signal throughout the volume, improves convergence at larger heights where the field is weaker, and further discourages trivial near-zero solutions. Note that the weighting parameters are part of the computational graph.

The full optimization objective is then
\begin{equation}
    \label{eq:combined-loss}
    \begin{aligned}
    \mathcal{L}_{\rm total} ={}& \lambda_{\rm bottom}
    \left\langle \mathcal{L}_{\rm bottom}\right\rangle \\
    &+ \lambda_{\rm potential-boundary}
    \left\langle \mathcal{L}_{\rm potential-boundary}\right\rangle \\
    &+ \lambda_{\rm height}
    \left\langle \mathcal{L}_{\rm height}\right\rangle \\
    &+ \lambda_{\rm ff}
    \left\langle w(z)\,\mathcal{L}_{\rm force-free}\right\rangle \\
    &+ \lambda_{\rm current}
    \left\langle w(z)\,\mathcal{L}_{\rm current}\right\rangle \, ,
    \end{aligned}
\end{equation}
where $\langle\cdot\rangle$ denotes the mean over the sampled points used to evaluate the corresponding loss term. As the default setting, we use $w_{\rm los}=0.7$, $\lambda_{\rm bottom}=1$, $\lambda_{\rm ff}=10^{-4}$, $\lambda_{\rm current}=10^{-4}$, $\lambda_{\rm height}=10^{-5}$, and $\lambda_{\rm potential-boundary}=10$. The $\lambda_{\rm current}$ weight is set to zero after the first 5,000 optimization steps. For the observational AR 13392 extrapolations, we use $\lambda_{\rm ff}=5\times10^{-4}$, as discussed in Appendix~\ref{sec:parameters}.

In contrast to our previous implementations, we evaluate the physics losses only on the randomly sampled volume points and exclude the potential field boundary as well as the lower and internal observational constraints from these terms. This avoids overly strict enforcement of the force-free condition at boundaries where the magnetic field is generally not expected to be force-free, and thereby improves the trade-off between the physics constraints and the imposed data conditions.

As the model, we use a basic multi-layer perceptron with sine activation functions, 256 neurons per layer, and 8 layers. In addition, we employ positional encoding to map the input coordinates into Fourier features as described by \citet{tancik2020fourier}. For this, we select 64 random frequencies per coordinate from a normal distribution with mean 0 and standard deviation $\sigma$ and encode the input coordinates 
\begin{align}
    \vec{v}_{\rm encoded} = \big[
        &\sin{(f_{x, 0} \cdot x)}, \cos{(f_{x, 0} \cdot x)}, \sin{(f_{x, 1} \cdot x)}, \ldots \notag \\
        &\ldots, \sin{(f_{z, 63} \cdot z)}, \cos{(f_{z, 63} \cdot z)},  x, y, z
    \big] \,,
\end{align}
where $f_{i,j}$ refers to the $j$th sampled frequency of coordinate $i$. This enables the neural network to efficiently learn high-frequency features. The encoding of the coordinate space into sine and cosine functions is a critical component for the spatial smoothness of the solution. Frequencies that are too small can result in smooth solutions and suppress high-frequency features, whereas higher frequencies can lead to artifacts and noisy inversions \citep[cf.][]{tancik2020fourier}. The primary tuning parameter for spatial resolution is $\sigma$, which determines the frequency bias of the neural network. Here, we select $\sigma=2$, which provides sufficient spatial detail while remaining stable during training. In principle, a SIREN implementation would also be possible, with the primary tuning parameter being the weighting factor of the initial layer; however, we found similar results and easier tuning with the positional encoding implementation.
The positional encoding increases the model's ability to represent small-scale spatial structure. This only leads to stable solutions when combined with the height-dependent normalization of the physics losses; without this normalization, the additional high-frequency flexibility tends to be used to form artificial near-zero-field regions separated by sharp transition layers.

For the setup with four A100 GPUs, we sample 32,768 coordinate points from the lower boundary and internal observational data, 65,536 randomly sampled coordinate points in the volume, and 8,192 coordinate points from the potential field boundary. Optimization is performed iteratively for 100,000 steps until convergence is reached, which yields stable solutions for all extrapolations. We use the Adam optimizer with an initial learning rate of $5\times10^{-4}$, which is exponentially decayed to $5\times10^{-5}$ over $10^5$ iterations.
These sampling numbers were chosen to fully utilize the available GPUs, although smaller sampling configurations also lead to stable convergence. In particular, when fewer GPUs are used, the number of samples is reduced linearly. The resulting extrapolations are typically similar, but require more compute time.

\subsection{Metrics}

To characterize magnetic connectivity, we use two field-line-based metrics: the squashing factor $Q$ and the twist $T$ \citep{Titov2002, Titov2007}.
The squashing factor is derived from the differential mapping of magnetic field lines. For a field line connecting two surfaces, $S_1$ and $S_2$, with on-surface coordinates $[x_1, y_1]$ and $[x_2, y_2]$, the Jacobian of the mapping is
\begin{equation}
D_{12}=\left(
\begin{array}{cc}
\frac{\partial x_2}{\partial x_1} & \frac{\partial x_2}{\partial y_1} \\
\frac{\partial y_2}{\partial x_1} & \frac{\partial y_2}{\partial y_1} \\
\end{array}
\right)
\equiv
\left(
\begin{array}{cc}
a & b \\
c & d \\
\end{array}
\right)
\label{eq:D+-}
\end{equation}
and the squashing factor at $[x_1,y_1]$ is
\begin{equation}
Q\left(x_1,y_1\right)=\frac{a^2+b^2+c^2+d^2}
{\left|\det D_{12}\right|}.
\label{eq:Q+-}
\end{equation}
A large $Q$ indicates strong local gradients in magnetic connectivity.
The twist $T$ quantifies the winding of magnetic field lines and is defined as
\begin{equation}
T=\int_L^{}\frac{\nabla\times\vec{B}\cdot\vec{B}}{4\pi B^2}\textrm{d}l,
\label{eq:tw}
\end{equation}
where $L$ denotes a magnetic field-line segment.
We compute both $Q$ and $T$ with FastQSL \citep{zhang2022fastqsl}, which uses graphics processing units (GPUs) for efficient evaluation over the full three-dimensional volume.

To quantify the amount of magnetic flux associated with the twisted structure, we define a mask $\mathcal{M}_h$ on a horizontal slice at height $h$ using the $T=0.5$ contour. The mask area is
\begin{equation}
    A = \sum_{i \in \mathcal{M}_h} \Delta A_i \, ,
\end{equation}
where $\Delta A_i$ is the area of pixel $i$. The signed and unsigned vertical magnetic fluxes through the same mask are then computed as
\begin{equation}
    \begin{aligned}
    \phi &= \sum_{i \in \mathcal{M}_h} B_z(x_i,y_i,h)\,\Delta A_i \, , \\
    |\phi| &= \sum_{i \in \mathcal{M}_h} \left|B_z(x_i,y_i,h)\right|\,\Delta A_i \, .
    \end{aligned}
\end{equation}
Here, $\phi$ gives the net flux and therefore includes cancellation between opposite polarities, whereas $|\phi|$ gives the total unsigned flux contained in the twist-mask region.

As a primary diagnostic for identifying concave-up field-line curvature, we evaluate the field-aligned gradient of the vertical magnetic field direction,
\begin{equation}
    \hat{\mathbf{B}} \cdot \nabla \hat{B}_z \, ,
\end{equation}
where $\hat{\mathbf{B}}=\mathbf{B}/|\mathbf{B}|$ and $\hat{B}_z$ denotes the vertical component of the unit magnetic field. Positive values indicate that the field line bends upward along the field direction, corresponding to the three-dimensional analogue of a bald-patch configuration. In combination with the squashing factor, this metric can therefore be used to identify the bald-patch separatrix surface \citep{1999A&A...351..707T}, which separates the flux-rope field from the underlying arcade field lines.

\begin{figure*}[ht]
    \centering
    \includegraphics[width=\textwidth]{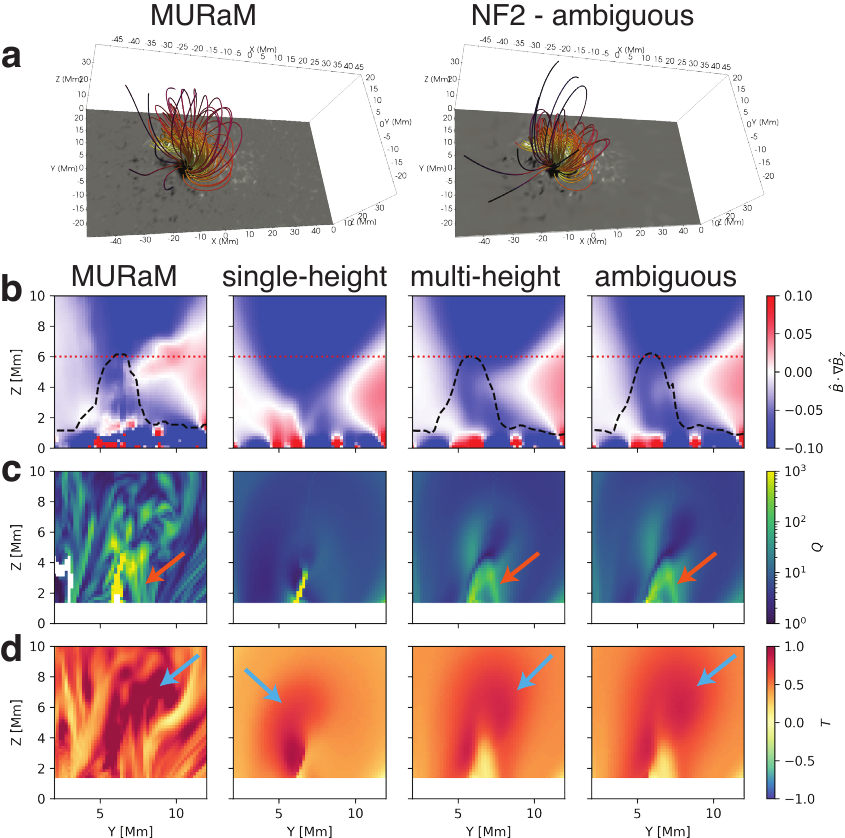}
    \caption{Magnetic flux rope benchmark. Panel (a) compares field-line renderings from the MURaM reference and NF2 extrapolations. Panels (b)--(d) show vertical slices at $x=-13$~Mm of field-line curvature, squashing factor, and twist, respectively, for the reference and extrapolation configurations. Multi-height constraints recover the separatrix layer at the flux-rope footpoints (red arrows) and the enhanced twist at greater heights (blue arrows) more accurately than the single-height extrapolation. The black dashed curves in panel (b) indicate the height surfaces of the $\tau=10^{-6}$ layer in the ground-truth MURaM atmosphere and recovered by the multi-height extrapolations. The red dotted line marks the approximate height of the field-line curvature reversal in the MURaM reference.}
    \label{fig:mfr}
\end{figure*}

\section{Data}

\subsection{MURaM simulation data}
\label{sec:data_simulations}

To evaluate the reliability of the extrapolated magnetic configuration in a controlled setting, we use two snapshots from the radiative-MHD simulations of \citet{rempel2023ApJmfr_sma}, computed with the Max Planck Institute/University of Chicago Radiative MHD code \citep[MURaM][]{rempel2017muram}. These simulations extend from the photosphere into the corona and model collisional sunspot configurations that produce filament-supporting magnetic structures. We select one snapshot representative of a magnetic flux rope (MFR; iteration 474000; Fig.~\ref{fig:mfr}) and one representative of a sheared magnetic arcade (SMA; iteration 399000; Fig.~\ref{fig:sma}). The two cases are particularly useful for our purposes because they provide distinct pre-eruptive magnetic configurations within a realistic thermodynamic and radiative-MHD atmosphere.

For each snapshot, we extract synthetic vector-magnetic observables on surfaces of constant optical depth computed at 500~nm. The $\tau=1$ layer is used as the photospheric constraint, while the $\tau=10^{-6}$ layer is used as the chromospheric constraint. In the extrapolation setup, the photospheric surface is assigned to $z=0$, although the actual $\tau=1$ surface in the MURaM atmosphere is geometrically corrugated and spans several hundred kilometers, especially in strong-field regions because of Wilson depression \citep{2003A&ARv..11..153S}. The chromospheric $\tau=10^{-6}$ surface exhibits substantially larger height variations, reaching multiple Mm. We retain these corrugated surfaces in the synthetic observables to mimic the challenges of real magnetograms. To emulate ambiguous magnetograms, we fold the transverse-field azimuth into the $0^\circ$--$180^\circ$ range.

For validation, we use the three-dimensional MURaM magnetic field as the reference solution. The comparison volume extends from the minimum height of the $\tau=1$ surface in the domain to 50\,Mm above it. This choice ensures that the full extrapolation volume is covered by the reference field, but it also means that parts of the MURaM reference start slightly below the nominal $z=0$ boundary used for the extrapolation. No additional artificial degradation, such as photon noise or spatial-resolution degradation, is applied. The synthetic experiment is therefore designed to isolate the dominant methodological challenges of this study: the mismatch between the lower-atmosphere magnetic field and the force-free assumption, the geometric corrugation of the observing surfaces, and the 180$^\circ$ ambiguity of the transverse magnetic field.

\subsection{SST observation data}
\label{sec:data}

For the observational application, we use spectropolarimetric data of NOAA AR 13392 (SHARP 9875) obtained with the Swedish Solar Telescope (SST; \citealt{sst2003scharmer}) and the CRisp Imaging SpectroPolarimeter (CRISP; \citealt{sharmer2008CRISP}) on 2023 August 6. The observations consist of a rectangular mosaic of 4$\times$8 different pointings, starting at 08:51~UTC, with a duration of approximately 7\,min. CRISP sampled the full Stokes profiles ([$I$, $Q$, $U$, $V$]) with 14 and 21 wavelength positions in the \ion{Fe}{I} 6173\,\AA~and \ion{Ca}{II} 8542\,\AA~lines, respectively. The spectral sampling varied between 40--360(70--850)\,m\AA~in the \ion{Fe}{I}(\ion{Ca}{II}) lines, with finer sampling close to the line cores than in the wings. The mosaic was constructed by cross-correlation of continuum maps in the overlapping areas, and it was downsampled by a factor of two to reduce the data volume. The combined map has a size of $1466\times2426$ pixels and a spatial scale of approximately 0.09~Mm~pix$^{-1}$.

We corrected the reduced CRISP data in both filters for residual polarization crosstalk using standard approaches \citep[e.g.,][]{1992ApJ...398..359S}.
The CRISP data were further processed to remove fringe-like patterns from Stokes $Q$, $U$, and $V$. These fringes, with amplitudes of a few percent of the signal, can propagate into the inferred magnetic field components if left untreated. Fringe removal in CRISP data is typically performed using a combination of two-dimensional Fourier filtering and principal component analysis to address the different spatial frequencies of the patterns \citep[e.g.,][]{2020A&A...644A..43P}. 

Given the large volume of data, we inferred the magnetic field components at different heights using two fast inversion codes based on widely used approximations. For the photospheric line (\ion{Fe}{I}), we perform Milne-Eddington (ME) inversions with the \texttt{PyMilne} code\footnote{\url{https://github.com/jaimedelacruz/pyMilne}} \citep{2019A&A...631A.153D}, which is based on analytical response functions \citep{2007A&A...462.1137O}. The line fits are generally of high quality, except in localized regions with pronounced Stokes-profile asymmetries that indicate velocity gradients beyond the scope of the ME approximation. Still, we expect the inferred magnetic field components to represent reasonable height-averaged quantities \citep[e.g.,][]{2010A&A...518A...2O}. 

For the chromospheric \ion{Ca}{II} line, we use the spatially coupled weak-field approximation (WFA) code\footnote{\url{https://github.com/morosinroberta/spatial_WFA}} developed by \citet{2020A&A...642A.210M}. We expect the WFA to be valid across most of the FOV, except possibly in the strongest magnetic concentrations within the sunspot umbra ($\gtrsim 1.5$\,kG), where the Zeeman broadening increases relative to the Doppler broadening. This limitation is therefore expected to have only a limited impact on the overall field extrapolation. In addition, the learned height mapping can absorb part of the resulting mismatch by assigning the chromospheric constraint to a height at which the extrapolated field strength is more consistent with the WFA-inferred field.

Because the observations were obtained sufficiently close to disk center (central helioprojective coordinates of approximately $T_x=302.7\arcsec$ and $T_y=70.0\arcsec$), we assume that projection effects are secondary. We therefore interpret the measured magnetic field in a local Cartesian reference frame, with the line-of-sight component corresponding to the $z$-axis.

\subsection{Context imaging from SDO/AIA and KSO}

To provide the large-scale context for the SST target region, we use co-temporal observations from the Atmospheric Imaging Assembly (AIA; \citealt{lemen2012aia}) and H$\alpha$ images from the Kanzelhöhe Observatory (KSO; \citealt{poetzi2021}). The AIA EUV channels trace the coronal and transition-region response of the filament system, while the KSO H$\alpha$ observations show the chromospheric filament morphology and its subsequent breakup. These context data are used to identify the eruptive filament segment, relate the SST field of view to the surrounding active region structure, and assess the subsequent reconfiguration of the filament system.

\section{Results}

Numerical setups can realize both flux-rope and arcade-type configurations \citep{rempel2023ApJmfr_sma}. While this provides direct insight into MHD processes in a realistic solar setting, the key question remains which magnetic configuration best reflects the physical reality of the solar atmosphere. To identify realistic magnetic configurations, we use data-driven magnetic field extrapolations.
To this end, we pursue two tasks. First, we validate our extrapolation method's ability to distinguish between the two magnetic configurations using realistic synthetic observations from MURaM simulations. We quantify performance against the known ground truth and identify the primary components that characterize different magnetic configurations (Sect. \ref{sec:simulations}). Second, we apply our approach to photospheric and chromospheric magnetograms of NOAA AR 13392 to determine the magnetic field configuration of a pre-eruptive filament (Sect. \ref{sec:observations}).

For the topological assessment, we use three complementary metrics: field-line curvature, squashing factor $Q$, and twist $T$. The curvature metric, $\hat{\mathbf{B}} \cdot \nabla \hat{B}_z$, identifies regions with upward (positive) or downward (negative) bending field lines and generalizes bald-patch-style diagnostics to the full three-dimensional case. In particular, flux rope configurations are characterized by curvature reversals, where upward or weakly curved field lines lie beneath downward-curved overlying fields. The squashing factor $Q$ highlights strong gradients in magnetic connectivity (quasi-separatrix layers), while $T$ quantifies the number of turns field lines wind about a common axis. Both $Q$ and $T$ are field-line-integrated quantities; therefore, even when viewing only slices, they provide an assessment of the three-dimensional topology. Together, these metrics provide a robust set for identifying distinct features of flux-rope and sheared arcade configurations. Since dipped fields can also occur in sheared arcades, we identify MFRs from the combined occurrence of curvature reversals, high-$Q$ separatrix structure, and a coherent enhancement in twist.

\subsection{Reliability of Estimated Configurations}
\label{sec:simulations}

\begin{figure}[ht]
    \centering
    \includegraphics[width=\linewidth]{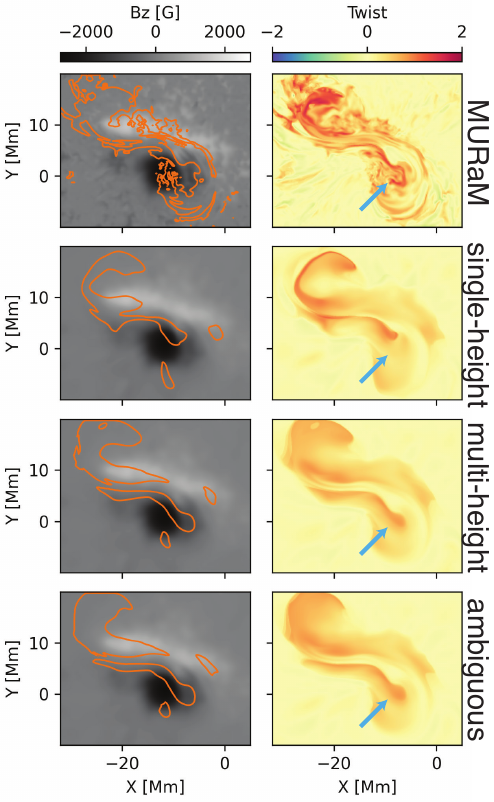}
    \caption{Twist map comparison for the MFR benchmark. The panels compare the MURaM reference with NF2 extrapolations using single-height, multi-height, and ambiguous multi-height constraints. Left panels show $B_z$ at a height of 1.2 Mm; right panels show the corresponding twist maps from the same horizontal slice. The single-height extrapolation recovers the central twist enhancement, but fails to recover the full spatial extent (blue arrow) and total twist of the structure.}
    \label{fig:twist_map}
\end{figure}

\begin{figure*}[t]
    \centering
    \includegraphics[width=\textwidth]{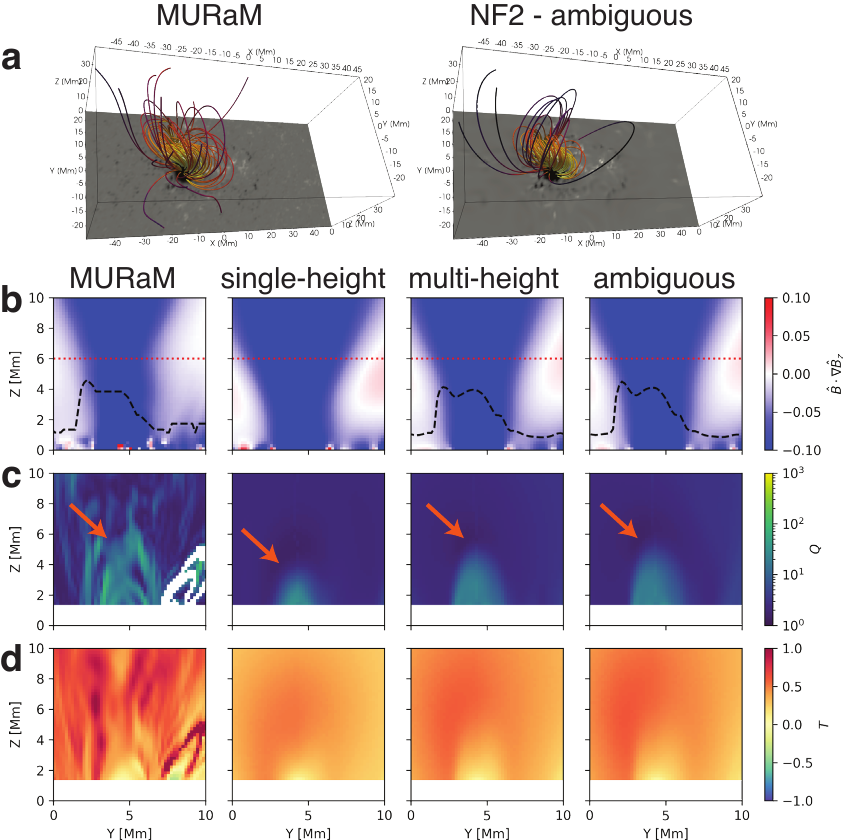}
    \caption{Sheared magnetic arcade benchmark. Panel (a) compares field-line renderings from the MURaM reference and NF2 extrapolations. Panels (b)--(d) show vertical slices at $x=-13$~Mm of field-line curvature, squashing factor, and twist, respectively, for the reference and extrapolation configurations. The SMA configuration is characterized by predominantly negative curvature, weakly diverging connectivity structure, and lower twist than the MFR case. Multi-height constraints improve the vertical placement of these signatures relative to the single-height extrapolation (red arrows). The black dashed curves in panel (b) indicate the height surfaces of the $\tau=10^{-6}$ layer in the ground-truth MURaM atmosphere and recovered by the multi-height extrapolations. The red dotted line marks the approximate height of the field-line curvature reversal in the MFR reference and is shown here for comparison.}
    \label{fig:sma}
\end{figure*}

We use the two MURaM snapshots described in Sect.~\ref{sec:data_simulations} to test whether NLFF extrapolations can distinguish between a pre-eruptive sheared magnetic arcade and a magnetic flux rope. The simulations provide an ideal benchmark because the magnetic configuration is known from the full three-dimensional reference field, while the extrapolations are constrained only by synthetic photospheric and chromospheric magnetograms.

For each MURaM snapshot, we perform three extrapolation experiments:
\begin{enumerate}
    \item \textit{single-height}: using only the disambiguated photospheric vector magnetogram;
    \item \textit{multi-height}: using disambiguated photospheric and chromospheric vector magnetograms with intrinsic height remapping \citep[cf.][]{jarolim2024multi_height};
    \item \textit{ambiguous}: using ambiguous photospheric and chromospheric vector magnetograms with intrinsic height remapping and intrinsic disambiguation.
\end{enumerate}
The ambiguous case maps the transverse magnetic field azimuth to the $0$--$180^\circ$ range before training and therefore tests whether the model can recover the correct vector orientation as part of the extrapolation. In this way, the three experiments progressively add the key observational complications addressed in this work: multi-height formation, corrugated observing surfaces, and azimuthal ambiguity.

For all extrapolations, we use the vector-potential formulation, $\mathbf{B}=\nabla\times\mathbf{A}$, which enforces $\nabla\cdot\mathbf{B}=0$ up to numerical precision. The force-free loss weight is set to $\lambda_{\rm ff}=10^{-4}$. Larger values yield similar magnetic configurations, but the lower weight provides a better match to the boundary constraints while maintaining stable force-free metrics. The remaining optimization settings follow the procedure described in Sect.~\ref{sec:training}, including height remapping, intrinsic disambiguation where applicable, and the improved convergence strategy.

We compare the extrapolations against the MURaM reference field in the volume from the minimum $\tau=1$ height to 50\,Mm. Topological diagnostics are evaluated above 1.5\,Mm, because the strong gradients and non-force-free structure of the MURaM simulation in the photosphere make field-line-based quantities computationally expensive and less interpretable at lower heights. Although the synthetic data would allow additional intermediate-height constraints, we restrict the experiments to photospheric and chromospheric layers to match the observational setup used in Sect.~\ref{sec:observations}.

Figure \ref{fig:mfr} summarizes the magnetic field extrapolation results for the magnetic flux rope configuration. Panel a presents an overview of the simulation volume (up to 40 Mm in height), together with field-line traces from the ground-truth MURaM magnetic field (left) and the NF2 multi-height extrapolation using ambiguous input vectors (right). Both solutions show a similar central structure with strongly twisted field lines, indicating good agreement in the reconstructed magnetic configuration. Overall, the extrapolated structure appears slightly more contracted, which is likely related to differences in boundary treatment (MURaM uses periodic boundaries, whereas NF2 extrapolations use potential field boundaries) and may also reflect missing plasma contributions in the extrapolations.

In Fig. \ref{fig:mfr}b--d, we show slices at $x=-13$ Mm of field-line curvature, squashing factor, and twist for the ground-truth MURaM data and the three extrapolation configurations. 
Because panel b shows only a single vertical cross-section, the curvature diagnostic should be interpreted as a local slice; additional dipped field lines are present along the extended rope axis but are not fully captured in this cut. The MURaM reference shows the expected flux-rope signatures: a transition from positive to strongly negative curvature (panel b), enhanced connectivity gradients in the central structure (panel c), and strongly twisted fields (panel d). In the single-height extrapolation, the transition to strongly downward-curved fields (i.e., the flux-rope core) is recovered but shifted by roughly 2 Mm toward lower heights. Consistently, the enhanced squashing-factor and twist signatures are also displaced downward, indicating a lower-lying reconstructed flux rope.

The multi-height extrapolations provide a systematic improvement. The curvature reversal is recovered at approximately the reference height (about 6 Mm), and the lower part of the rope is matched more accurately. In panel b, we also show the inferred $\tau=10^{-6}$ height as a dashed line. The close agreement with the reference surface demonstrates that the model can intrinsically recover formation-height geometry as part of the extrapolation. This comparison also highlights where chromospheric constraints add the most information, as reflected by improved structural agreement near the remapped height surface. In particular, regions with dense plasma structure benefit strongly because higher-lying $\tau$ surfaces provide significant additional constraints. In contrast, in adjacent strong-field regions where the chromospheric formation height is around $\sim1$ Mm, the multi-height and single-height extrapolations yield similar results.

The squashing-factor maps further resolve the separatrix structure between the weakly twisted underlying arcade and the highly twisted flux-rope core. The loop-like high-$Q$ pattern is characteristic of a hyperbolic flux-rope configuration and is clearly visible in the MURaM reference and the multi-height solutions, but less distinct in the single-height extrapolation. The flux rope is located near the top of this loop-like separatrix, as confirmed by the twist maps. The multi-height extrapolations better reproduce the MFR configuration, including the enhanced twist around $\sim7$ Mm and the lower-twist arcade footpoints. In contrast, the single-height extrapolation captures the overall magnetic structure but places the twisted core less accurately. All extrapolations remain smoother than the MURaM reference and show reduced fine structure and lower twist amplitudes, which is expected given the NLFF approximation and the absence of explicit plasma dynamics.

Figure~\ref{fig:twist_map} shows horizontal slices of the magnetic twist map for the MURaM reference field and the extrapolated fields. The twist is evaluated from a height of $h=1.152$~Mm. This is the lowest height at which the field-line tracing remains robust and does not become trapped in the highly structured photospheric field of the MURaM simulation. The left panels show $B_z$ at this height, with the $T=0.5$ contour overlaid, while the right panels show the corresponding twist maps. The multi-height extrapolations improve the recovered spatial extent of the modeled flux rope, although the extrapolations remain smoother and less twisted than the MURaM reference.

The twist-mask and flux analysis confirms this trend quantitatively (Table~\ref{tab:mask_flux_summary}). The multi-height extrapolations recover a larger twist-mask area and unsigned flux than the single-height extrapolation, increasing $|\phi|$ from $4.72\times10^{20}$~Mx to $7.29$--$8.11\times10^{20}$~Mx and moving closer to the MURaM value of $12.4\times10^{20}$~Mx. The signed flux remains much smaller than the unsigned flux in all cases, indicating the expected cancellation between opposite polarities within the twisted region.
We use $T = 0.5$ as a practical mask for the coherent twisted channel rather than as a universal instability or flux-rope threshold. Tests with nearby thresholds yield the same qualitative results between the single-height and multi-height extrapolations.

\begin{table}
\centering
\caption{Area and magnetic flux of the twist-mask region for the MURaM reference field and NF2 extrapolations. The table lists the mask area $A$, signed magnetic flux $\phi$, and unsigned magnetic flux $|\phi|$ evaluated at the horizontal slice height $h=1.152$~Mm.}
\label{tab:mask_flux_summary}
\begin{tabular}{lrrr}
\hline
Dataset & A [Mm$^2$] & $\phi$ [$10^{20}$ Mx] & $|\phi|$ [$10^{20}$ Mx] \\
\hline
MURaM          & 249.90 & $-0.47$ & $12.4$ \\
single        & 114.61 & $-0.28$ & $4.72$ \\
multi       & 182.48 & $-0.25$ & $7.29$ \\
amb.  & 185.57 & $-0.26$ & $8.11$ \\
\hline
\end{tabular}
\end{table}

Figure \ref{fig:sma} shows the corresponding analysis for the sheared magnetic arcade case. In panel a, the extrapolations recover a magnetic structure that is again similar to the MURaM reference but slightly more contracted. The visual similarity between field-line renderings in the MFR (Fig. \ref{fig:mfr}) and SMA (Fig. \ref{fig:sma}) cases also illustrates that field-line plots alone are insufficient to distinguish between the two magnetic configurations.

The MURaM reference exhibits the expected SMA signatures: predominantly negative curvature (downward-curved fields; panel b), weakly diverging field lines with maximum divergence near the arcade center (panel c), and overall less twisted fields (panel d). Both single-height and multi-height extrapolations recover these signatures and reproduce the reference SMA-like field structure. Similar to the MFR case, the single-height solution places the structure slightly too low (red arrows), whereas multi-height constraints improve the vertical placement.

Overall, extrapolations with intrinsic magnetic field disambiguation yield results that are equivalent to those obtained from disambiguated inputs. This behavior is critical for observational applications, where azimuth ambiguities are unavoidable and are typically difficult to resolve for corrugated formation-height surfaces.

In summary, the MFR classification is identified by the appearance of (i) a curvature reversal from positive to strongly negative values marking a dipped flux-rope core, (ii) enhanced connectivity gradients that outline a separatrix-like boundary around that core, and (iii) a concentrated high-twist channel at corresponding heights. In contrast, the SMA case shows predominantly negative curvature without a compact twisted core, weaker and more distributed connectivity gradients, and overall lower twist.

\begin{figure}[t]
    \centering
    \includegraphics[width=\columnwidth]{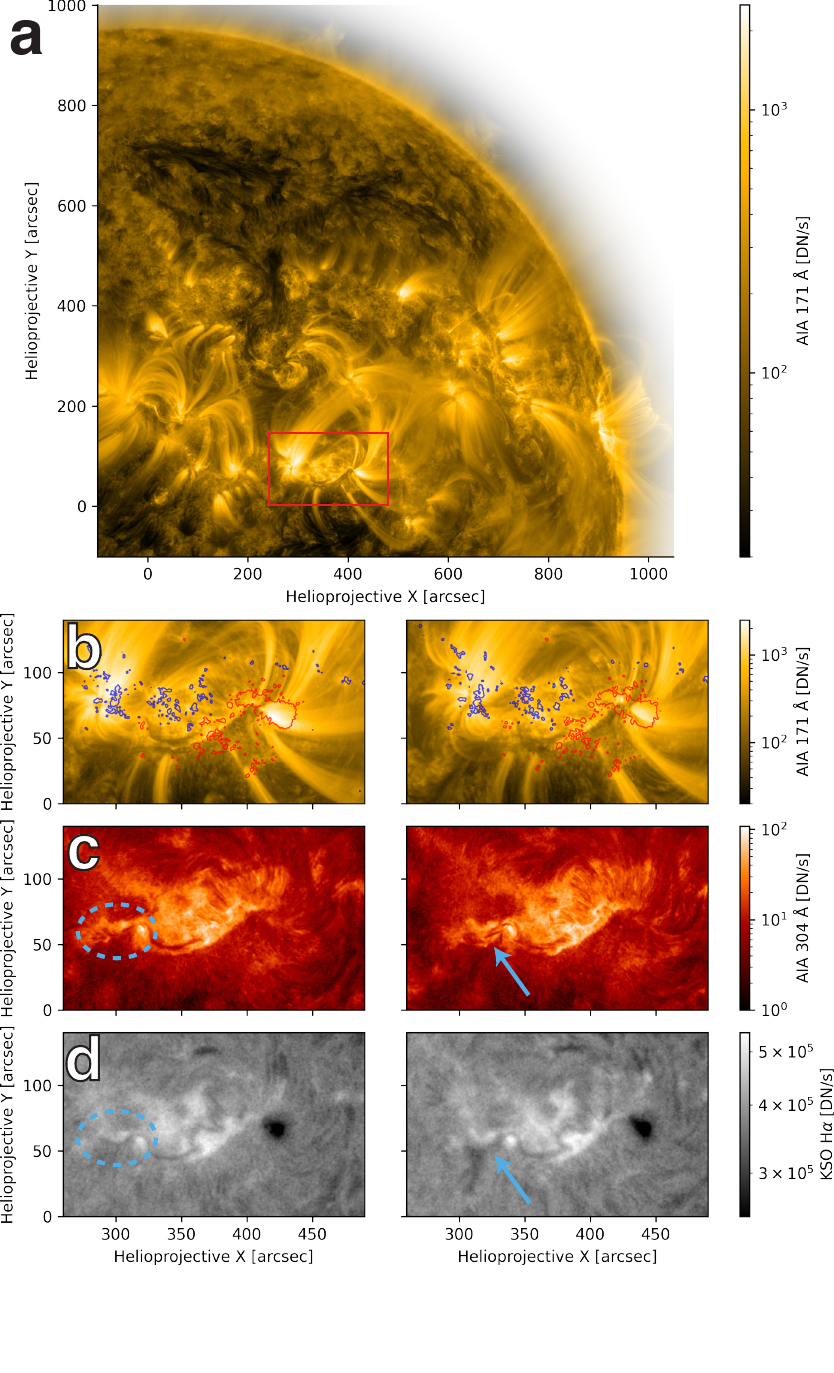}
    \caption{Observational context of the target filament in active region 13392 on 2023 August 6. Panels (a) and (b) show SDO/AIA 171~{\AA} EUV emission at 09:12~UT and 10:48~UT, respectively. Panel (c) shows SDO/AIA 304~{\AA} emission at 09:12~UT, and panel (d) shows the corresponding KSO H$\alpha$ observation. The left column is close in time to the SST mosaic, while the right column shows the same region 1~hr~36~min later. Magnetic field contours show the radial photospheric field at $-500$~G and $+500$~G, shown as blue and red contours, respectively. The blue ellipse marks the primary filament segment analyzed in this work, and the blue arrows highlight the region where the filament subsequently breaks up.}

    \label{fig:observations}
\end{figure}

\subsection{Magnetic Configuration of the Solar Filament}
\label{sec:observations}

We apply this framework to SST observations of AR 13392 from 2023 August 6 at 08:51 UTC. Figure \ref{fig:observations} provides an overview of the target region in Extreme Ultraviolet (EUV) emission, combining co-temporal observations from the SDO/AIA and H$\alpha$ observations from KSO. The left column shows the closest available context images to the SST mosaic, while the right column shows the same region 1~hr~36~min later, at 10:48~UTC, when the filament channel exhibits clear magnetic reconfiguration. Our analysis focuses on the left segment of the filament (blue circle), which appears unstable compared to the right segment. Both AIA 304~{\AA} and KSO H$\alpha$ observations show a filament break-up in this region (blue arrows).

We use co-spatial and co-temporal photospheric and chromospheric vector magnetograms of the full FOV, based on the processed and mosaiced data described in Sect. \ref{sec:data}. We perform both single- and multi-height extrapolations. For these extrapolations, we increase the force-free weight to $\lambda_{\rm ff}=5 \times 10^{-4}$, which provides a good compromise between boundary constraints and force-freeness according to the metrics in Appendix \ref{sec:parameters}. For the remaining optimization, we follow the same setup as for the synthetic data case, using the default parameters and training for 20 epochs until convergence is reached.

\begin{figure*}[t]
    \centering
    \includegraphics[width=\textwidth]{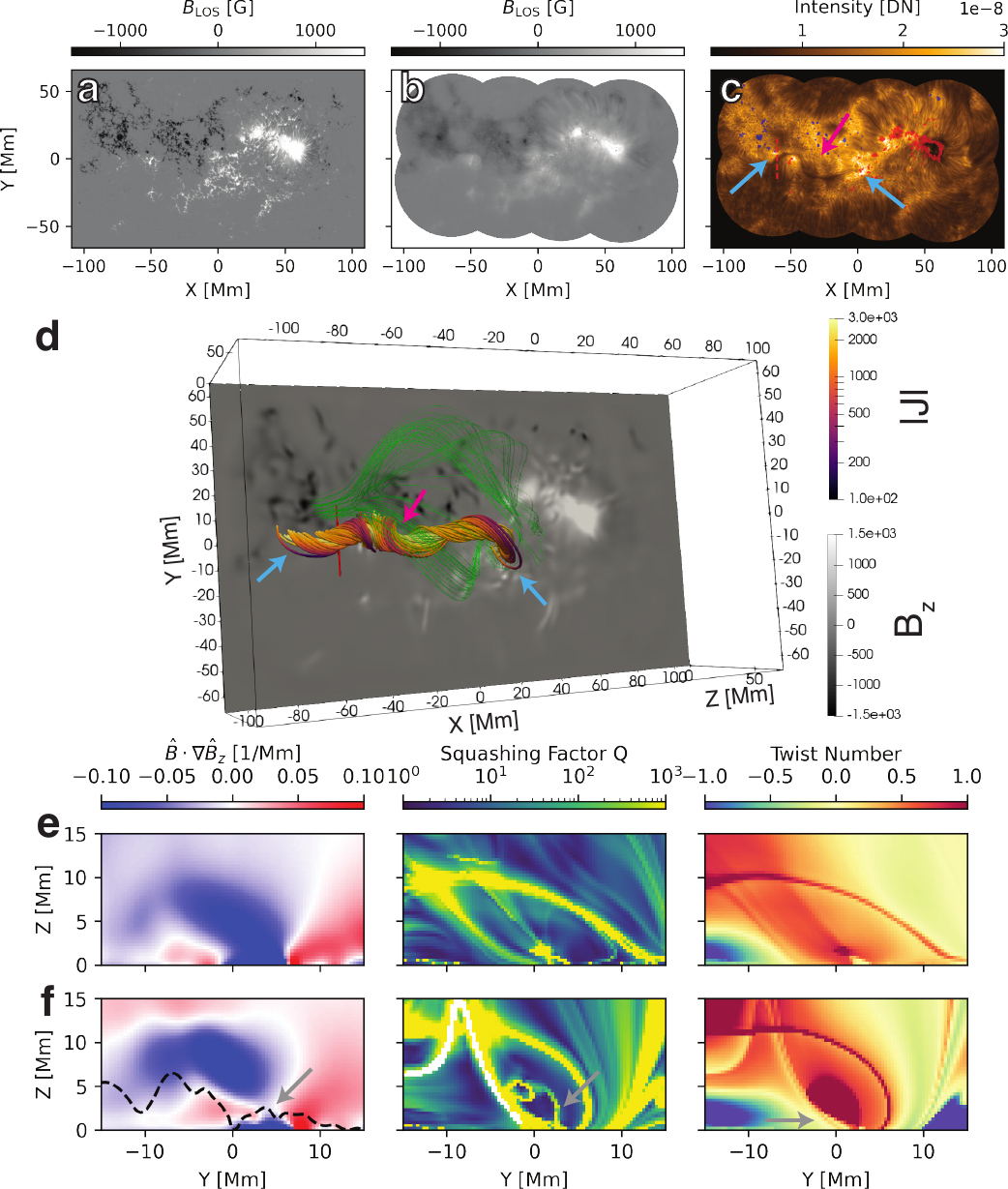}
    \caption{
        Topology analysis of the observed filament from SST-constrained NLFF extrapolations. Panels (a) and (b) show the photospheric and chromospheric line-of-sight magnetic field maps, respectively. Panel (c) shows the Ca II line-wing intensity image of the filament. Panel (d) shows magnetic field-line traces from the multi-height extrapolation, colored by $||J||$, together with the overlying strapping field (green). Panels (e) and (f) show slices at $x=-60$ Mm of field-line curvature, squashing factor, and twist for the single-height and multi-height solutions. The multi-height solution recovers clearer flux-rope signatures, including dipped field lines and strong connectivity gradients around the highly twisted field-line channel. The dashed black curve in panel (f) marks the inferred chromospheric height surface.}
    \label{fig:sst_result}
\end{figure*}

To evaluate the resulting magnetic field extrapolation, we compute the same metrics as in the synthetic case, identify the primary magnetic structures, and relate them to the simulation results. Figure \ref{fig:sst_result} shows the main results of this investigation. Panels (a) and (b) show the photospheric and chromospheric line-of-sight magnetic field, respectively. Panel (c) shows an intensity image of the \ion{Ca}{II} line wing, where the solar filament is clearly visible. The color contours show the corresponding magnetic field at $-$1500 G (blue) and +1500 G (red). 

Panel d shows field-line traces of the multi-height extrapolation, illustrating the magnetic structure (color-coded by absolute current density $||J||$) and the overlying field (green). Compared with the observed filament structure in panel c, the field-line structure shows strong agreement in overall morphology, with field lines rooted in the same positive and negative magnetic field regions (blue arrows). The pink arrow highlights the reference point for the magnetic field-line traces, where the extrapolation reproduces the footpoints of the filament and the sharp kink in orientation. The field-line traces indicate a two-part flux-rope system, rather than a single continuous filament. The green field-line traces are extracted from an apparent null-point region close to the central footpoint of the filament. The field-line connectivity shows that the overlying field largely spans the right (western) part of the filament structure, while the left part has significantly weaker strapping fields. This magnetic field configuration is consistent with the greater stability of the right structure, which is supported by the overlying field.

To determine the supporting structure of the filament, we compare slices at $x=-60$ Mm, as indicated by the red line in panels c and d, of field-line curvature, squashing factor, and twist, analogous to Sect. \ref{sec:simulations}. Panels e and f in Fig. \ref{fig:sst_result} correspond to the single-height and multi-height extrapolations, respectively. In the multi-height extrapolation, we identify distinct signatures of a flux rope: a reversal in field-line curvature, strong separatrix layers in the lower and upper atmosphere, and enhanced field-line twist at greater heights with distinctly lower twist near the footpoint. Together, these metrics strongly suggest that the solar filament is supported by a magnetic flux-rope configuration prior to eruption.

In comparison, the single-height extrapolation shows a less clear signature (panel e): no distinct change in field-line curvature, a more arcade-like structure, less distinct and lower-lying separatrix layers, and overall lower twist, although with similar large-scale morphology. Both the squashing-factor and twist maps still show flux-rope-like signatures, but they indicate a substantially lower-lying magnetic structure. Note that the squashing-factor and twist maps are field-line-integrated quantities, whereas the curvature map corresponds to a single slice. Therefore, although the curvature map locally suggests a sheared magnetic arcade, clearer flux-rope signatures are recovered along the filament channel. Specifically, close to the footpoint at $x\sim-50$ Mm, both curvature maps show a curvature reversal; however, for this evaluation, the most relevant region is the extended eruptive part of the filament. Overall, this evaluation is consistent with the MURaM-slice results: the height and morphology of the flux rope are better captured by the multi-height extrapolation, while both single- and multi-height extrapolations remain, in principle, capable of identifying the magnetic configuration.

Compared with the metrics in Sect. \ref{sec:simulations}, the extrapolations show increased complexity (e.g., strong separatrix layers above 10 Mm), which we interpret as interactions with the strapping magnetic field. The MURaM simulation represents a more isolated magnetic configuration, whereas the observational case contains a quadrupole-type configuration in which substantial changes in field-line connectivity at greater heights are expected between the central magnetic structure and the surrounding magnetic field.

The estimated height surface (dashed black line in Fig. \ref{fig:sst_result}f) further highlights the contribution of chromospheric constraints. The flux-rope center is mapped to $\sim3$ Mm and coincides with the curvature-reversal location, consistent with the inferred flux rope axis. We also find a pronounced height increase in the quiet-Sun region. In addition, some locations are mapped close to the photospheric level. These low-height solutions are mainly associated with weak magnetic field regions, where the chromospheric constraint provides only a weak optimization signal. In combination with the regularization toward minimal height offsets, mapping these locations to lower heights can therefore become the preferred solution. This behavior differs from the MURaM reference, where the recovered height surfaces show a smoother variation that more closely follows the expected $\tau$ surface. In the observational case, we attribute this deviation to a combination of effects. Potential non-physical contributors include (1) inversion assumptions, (2) neglected projection effects, and (3) different effective formation heights for horizontal and vertical field components. A potential physical contributor is optically dense plasma along the observed line of sight. Although these effects are not modeled explicitly here, the current setup still yields sufficiently high-quality extrapolations. Incorporating these additional terms is a natural next step for future work.

\section{Discussion}
\label{sec:discussion}

In this study, we demonstrated that NLFF extrapolations can distinguish between sheared magnetic arcade and magnetic flux rope configurations, as shown using synthetic data obtained from full radiative MHD simulations. The addition of chromospheric magnetic field observations significantly improves the robustness of the extrapolations and more accurately recovers the primary topological features used to identify magnetic flux ropes.

For the sheared magnetic arcade case, the extrapolations consistently recover the main topological features, whereas the magnetic flux rope case is more complex and shows differences between single- and multi-height extrapolations. In this case, the addition of chromospheric magnetic field constraints enables more robust identification of topological features, such as dipped field lines, the quasi-separatrix layer between the flux rope and the underlying arcade field, and improved agreement in the height and twist of the flux rope channel. The twist-mask flux analysis supports this conclusion quantitatively, showing that multi-height constraints recover a larger fraction of the MURaM unsigned twisted flux than the single-height extrapolation.
Overall, the results demonstrate that the 180$^\circ$ azimuthal ambiguity can be accounted for intrinsically within the model without trade-offs in extrapolation quality. Beyond the multi-height applications emphasized here, the same ambiguity-invariant formulation can also be applied to photosphere-only extrapolations, enabling automatic disambiguation of vector magnetograms within the extrapolation optimization itself. The automatic-disambiguation tests in Appendix~\ref{sec:auto_disambiguation} confirm this result for the known MURaM reference case and provide additional photospheric examples.

For the application to observations, we build on the advanced observing capabilities of SST/CRISP, using photospheric and chromospheric spectropolarimetric observations of NOAA AR 13392 from a mosaic encompassing the entire region. The extrapolations of the observational data show increased connectivity and topological complexity relative to the simulation. We focused our investigation on an elongated AR filament that later erupted. Observations of the filament structure in EUV, H$\alpha$, and \ion{Ca}{II} line wings show strong similarity with the extrapolation result in terms of extent, footpoint anchoring, and shape. Both single-height and multi-height extrapolations show highly twisted fields associated with the filament channel and signatures of an MFR configuration. The inclusion of chromospheric magnetograms results in more distinct dipped field lines and a twisted magnetic field channel at about 3 Mm, providing strong evidence that the filament is supported by a flux-rope-like magnetic configuration hours before the filament eruption.

The inferred height surface associated with the chromospheric magnetic field derived from \ion{Ca}{II} should be interpreted with caution. In the observational case, the recovered heights can be affected by inversion approximations, neglected projection effects, different effective formation heights for horizontal and vertical field components, and optically thick plasma along the line of sight.

A primary next step is to account for spherical geometry and projection effects, enabling magnetic field extrapolations at higher inclination angles. In particular, SOLIS/VSM \citep{2006ASPC..358...92H} provides daily photospheric and chromospheric magnetic field measurements that could support global multi-height extrapolations of the solar magnetic field. Such extrapolations could improve estimates of global magnetic connectivity and magnetic energy distribution, while also advancing our understanding of large-scale interactions such as sympathetic solar eruptions \citep{2025A&A...694A..74G}.

In this study, we find that the observed filament eruption was preceded by a magnetic configuration consistent with a flux-rope configuration, suggesting that the eruption is more consistent with destabilization of a pre-existing flux rope than with eruption-time flux-rope formation from a sheared arcade. This result demonstrates the diagnostic power of chromospheric vector-field constraints, but a single event cannot resolve whether filament eruptions are generally dominated by pre-existing flux ropes or whether sheared-arcade configurations also become eruptive under certain conditions. Addressing this question requires statistical applications of multi-height extrapolations to larger filament samples, in order to quantify the occurrence rates of different pre-eruptive magnetic configurations, identify their formation pathways, and follow their temporal evolution toward eruption.

Such studies depend on routine chromospheric magnetic field measurements. Ground-based solar observatories and instruments (e.g., SST/CRISP, SAMNet, and DKIST/ViSP; \citealt{sst2003scharmer, 2022JSWSC..12....2E, 2022SoPh..297...22D}) can already provide such observations, but adverse observing conditions and cadence limitations remain significant challenges. The new era of space-based spectropolarimetry will help fill this gap: with missions such as STRUVE \citep{Gamaunt2022STRUVE} and CMEx \citep{kalinowski2025cmex}, new chromospheric magnetic field data series will better constrain data-driven simulations of the solar corona and improve our understanding of magnetic reconfiguration prior to solar eruptions.

\section{Code Availability}
\label{sec:code_availability}

The code used in this study is published as open-access software. This work includes major improvements to the NF2 code base, and the full release associated with this manuscript (source code, experiment scripts, and figure-generation utilities) is provided alongside this publication. Public repositories and archival records are available at GitHub: \url{https://github.com/RobertJaro/NF2} and Zenodo: \citep{robert_jarolim_2026_20533552}. 
All extrapolation results used in this study are provided online via Globus: \url{https://app.globus.org/file-manager?origin_id=3f7d542e-a2b3-4a9a-9a0b-4c4d966c9601&origin_path=%2F}.

\begin{acknowledgments}

This work is based on observations obtained with the Swedish 1-m Solar Telescope (SST), operated on the island of La Palma by the Institute for Solar Physics of Stockholm University. We gratefully acknowledge that these observations were enabled through the SOLARNET Access Programme, under the sponsored project “Multi-height magnetic field measurements using a magneto-optical filter for neural network driven magnetic field simulations.” This research has received financial support from the European Union’s Horizon 2020 research and innovation programme under grant agreement No. 824135 (SOLARNET).
R.J. was supported by the NASA Jack-Eddy Fellowship. This material is based upon work supported by the NSF National Center for Atmospheric Research (NCAR), which is a major facility sponsored by the U.S. National Science Foundation under Cooperative Agreement No. 1852977. We would like to acknowledge high-performance computing support from the Derecho system (doi:10.5065/qx9a-pg09) provided by NCAR, sponsored by the National Science Foundation.
M.B.K. is grateful for the Leverhulme Trust Fund ECF-2023-271.  M.B.K. and R.E. acknowledge the NKFIH OTKA (Hungary, grant No. K142987) and the ESA COMMERCIAL APPLICATIONS OF SPACE WEATHER DATA FS - EXPRO+1-12676 2025.
R.E. acknowledges the Science and Technology Facilities Council (STFC, grant No. ST/M000826/1) UK, the NKFIH (OTKA, grant No. K142987) Hungary for enabling this research. R.E. is also grateful to PIFI (China, grant No. 2024PVA0043), and the NKFIH Excellence Grant TKP2021-NKTA-64 (Hungary).
D.K. acknowledges the Georgian Shota Rustaveli National Science Foundation project FR-22-7506.
The authors acknowledge the use of ChatGPT and Prism (OpenAI) for language editing and stylistic improvements. The authors are solely responsible for the scientific content, analysis, and conclusions presented in this work.

\end{acknowledgments}

\begin{contribution}


\end{contribution}

%
\facilities{SST/CRISP \citep{sst2003scharmer, sharmer2008CRISP}, SDO/HMI \citep{pesnell2012sdo, schou2012hmi, bobra2014sharps}, SDO/AIA \citep{lemen2012aia}, KSO \citep{poetzi2021}}

\software{Astropy \citep{2013A&A...558A..33A,2018AJ....156..123A,2022ApJ...935..167A},
          SunPy \citep{sunpycommunity2020, sunpysoftware2020},
          PyTorch \citep{pytorch2019_9015}, 
          spatial\_WFA \citep[]{2020A&A...642A.210M},
          PyMilne \citep[]{2019A&A...631A.153D}
          }


\appendix

\section{Automatic Disambiguation}
\label{sec:auto_disambiguation}

\begin{figure}[ht]
    \centering
    \includegraphics[width=\linewidth]{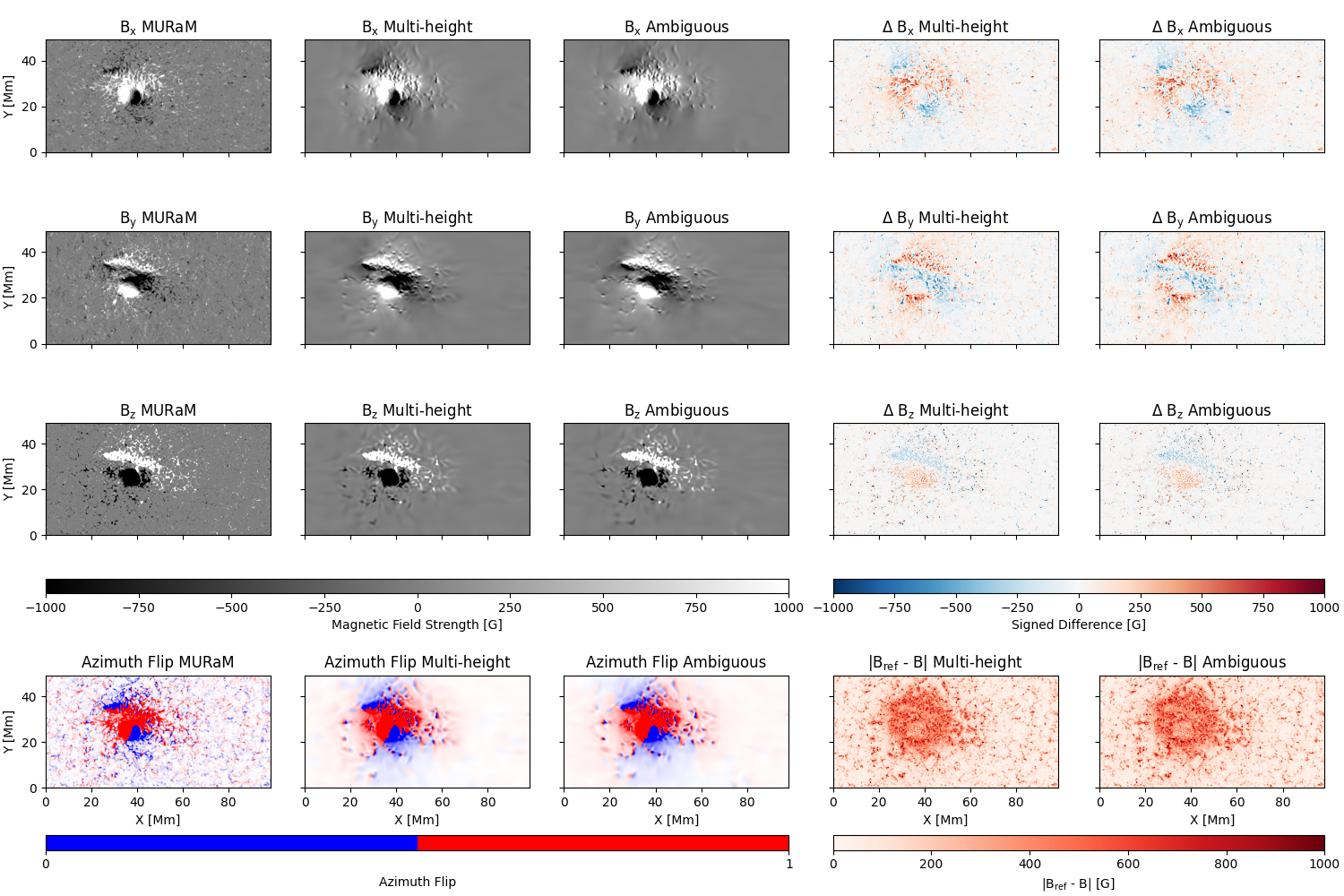}
    \caption{Automatic disambiguation benchmark for the MURaM synthetic magnetogram. The panels compare the reference vector magnetogram with boundary fields recovered from extrapolations using either disambiguated or ambiguous input data. Rows show the magnetic field components for the MURaM reference, the multi-height extrapolation, the ambiguous multi-height extrapolation, and the corresponding difference maps. The bottom row shows locations where the transverse field is flipped by $180^\circ$, with transparency scaled by the local field strength, together with the absolute vector magnetic field difference.}
    \label{fig:muram_disambiguation}
\end{figure}

\begin{figure}[ht]
    \centering
    \includegraphics[width=\linewidth]{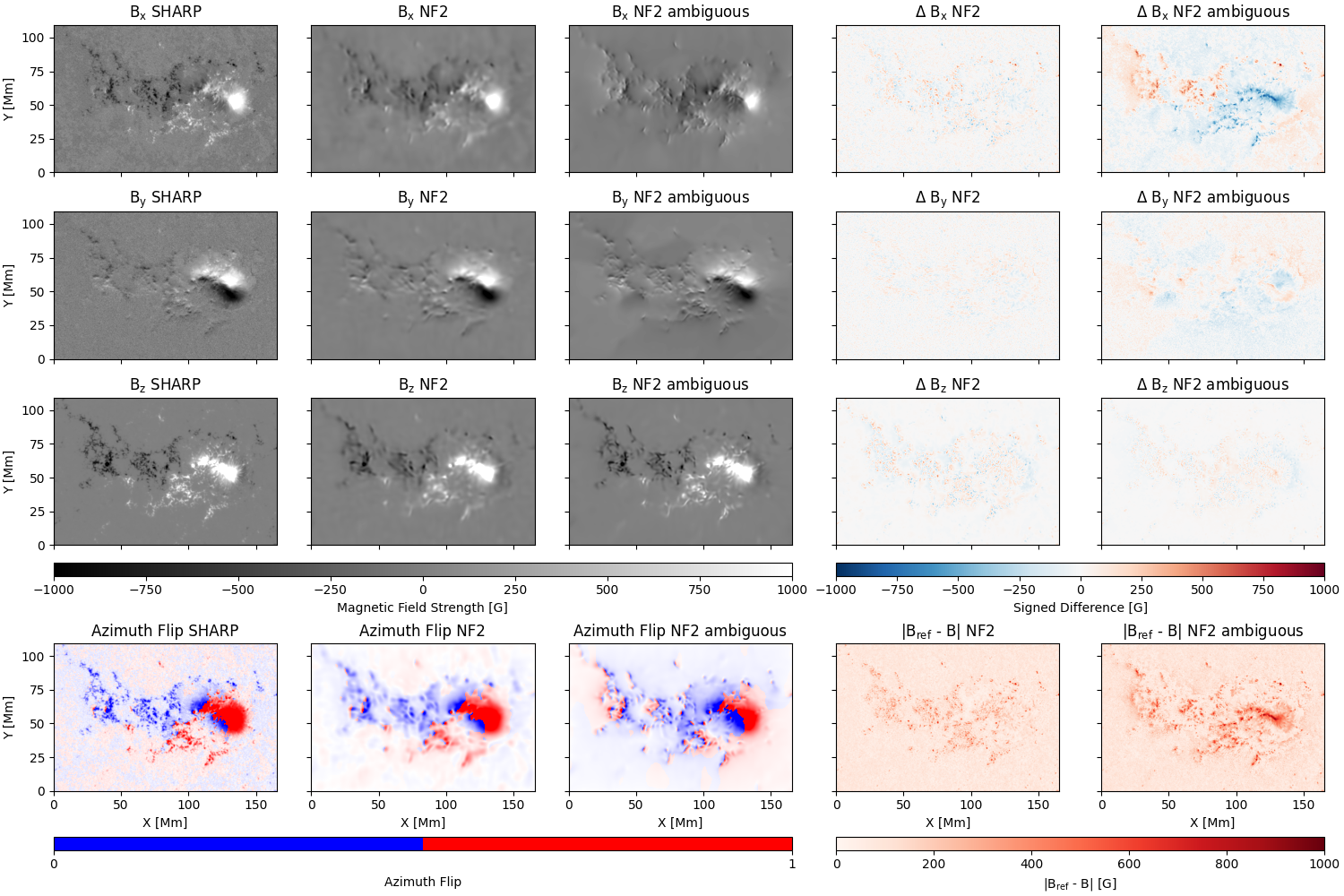}
    \caption{Automatic disambiguation comparison for the SHARP vector magnetogram, following the layout of Figure~\ref{fig:muram_disambiguation}. The panels compare the SHARP pipeline-disambiguated boundary field with the field recovered when the transverse-field ambiguity is resolved intrinsically during the NF2 extrapolation.}
    \label{fig:sharp_disambiguation}
\end{figure}

Figures~\ref{fig:muram_disambiguation} and \ref{fig:sharp_disambiguation} compare the boundary magnetic fields obtained with imposed and intrinsic azimuthal disambiguation. In the MURaM benchmark, extrapolations constrained by disambiguated and ambiguous vector fields recover nearly identical boundary fields. This demonstrates that the coupled optimization can recover the correct magnetic configuration directly from ambiguous transverse field measurements. Differences in weaker magnetic flux elements are present in both NF2 solutions, as expected from the trade-off between the force-free model assumption and the non-force-free magnetic structure of small-scale photospheric fields.

The comparison of the SHARP HMI vector magnetic field  data product \citep{bobra2014sharps} shows larger differences between the pipeline-disambiguated boundary field and the intrinsically ambiguity-resolved solution. When the pipeline disambiguation is imposed, the extrapolation follows the input magnetogram closely. When the ambiguity is optimized intrinsically, the solution permits larger-scale changes in the transverse-field orientation. These differences are accompanied by lower force-free metrics: the current-weighted angle decreases from $\theta_J=13.57^\circ$ to $9.42^\circ$, and $\sigma_J$ decreases from $0.235$ to $0.164$. The normalized divergence also decreases from $\langle|\nabla\cdot\mathbf{B}|/|\mathbf{B}|\rangle=1.53\times10^{-4}$ to $9.64\times10^{-5}$.

These results indicate that intrinsic disambiguation can reproduce the known MURaM reference solution without a measurable performance trade-off. For the SHARP case, the intrinsically disambiguated solution differs from the pipeline-disambiguated field and yields lower force-free and divergence metrics for this extrapolation setup. However, this should be interpreted as the ambiguity resolution that is most self-consistent with the adopted NLFF optimization objective, rather than as an independently validated ground-truth disambiguation. A more detailed comparison between existing disambiguation methods and our intrinsic disambiguation approach should be addressed in a separate study to quantify their differences and assess their impact on magnetic field extrapolation results.

\section{Parameter selection and performance variation}
\label{sec:parameters}

Table~\ref{tab:extrapolation_metrics} summarizes the numerical quality of the extrapolations. We compute the mean absolute divergence, $\langle|\nabla\cdot\mathbf{B}|\rangle$, and its field-strength-normalized form, $\langle|\nabla\cdot\mathbf{B}|/|\mathbf{B}|\rangle$, as measures of solenoidality. Force-freeness is quantified by the current-weighted angle $\theta_J$ between $\mathbf{J}=\nabla\times\mathbf{B}$ and $\mathbf{B}$, and by the corresponding current-weighted sine metric $\sigma_J$, with lower values indicating a more force-free solution. All extrapolations achieve low divergence and low $\theta_J$, indicating high-fidelity solutions that satisfy the force-free condition well. The ambiguous multi-height extrapolations achieve performance scores very similar to the disambiguated multi-height cases, demonstrating that the intrinsic disambiguation can operate on ambiguous vector fields without a measurable loss in extrapolation quality. For the observational AR 13392 case, $\lambda_{\rm ff}=5\times10^{-4}$ was adopted because it substantially improves the metrics relative to $10^{-4}$ (e.g., for the multi-height case, $\theta_J$ decreases from $9.39^\circ$ to $5.84^\circ$ and $\sigma_J$ from 0.163 to 0.102), while avoiding the stronger weighting of $10^{-3}$ that would place more emphasis on the force-free constraint relative to the observational boundary data.

\begin{table}[h]
\centering
\caption{Evaluation metrics for the MURaM and observational extrapolations.}
\label{tab:extrapolation_metrics}
\begin{tabular}{llcccc}
\hline
Dataset & Setup & $\langle|\nabla\cdot\mathbf{B}|\rangle$ 
& $\langle|\nabla\cdot\mathbf{B}|/|\mathbf{B}|\rangle$ 
& $\theta_J$ [deg] & $\sigma_J$ \\
\hline
MURaM MFR & single & $3.16\times10^{-2}$ & $1.01\times10^{-4}$ & 8.19 & 0.143 \\
MURaM MFR & multi & $2.39\times10^{-2}$ & $6.12\times10^{-5}$ & 6.37 & 0.111 \\
MURaM MFR & multi amb. & $2.36\times10^{-2}$ & $5.88\times10^{-5}$ & 6.56 & 0.114 \\
\hline
MURaM SMA & single & $1.64\times10^{-2}$ & $5.07\times10^{-5}$ & 10.48 & 0.182 \\
MURaM SMA & multi & $1.13\times10^{-2}$ & $3.07\times10^{-5}$ & 7.68 & 0.134 \\
MURaM SMA & multi amb. & $1.19\times10^{-2}$ & $2.59\times10^{-5}$ & 7.46 & 0.130 \\
\hline
AR 13392 & single, $\lambda_{\rm ff}=10^{-4}$ & $1.92\times10^{-2}$ & $7.18\times10^{-5}$ & 15.80 & 0.272 \\
AR 13392 & multi, $\lambda_{\rm ff}=10^{-4}$ & $1.37\times10^{-2}$ & $5.23\times10^{-5}$ & 9.39 & 0.163 \\
AR 13392 & single, $\lambda_{\rm ff}=5\times10^{-4}$ & $1.04\times10^{-2}$ & $3.70\times10^{-5}$ & 10.00 & 0.174 \\
AR 13392 & multi, $\lambda_{\rm ff}=5\times10^{-4}$ & $5.86\times10^{-3}$ & $2.36\times10^{-5}$ & 5.84 & 0.102 \\
AR 13392 & single, $\lambda_{\rm ff}=10^{-3}$ & $7.25\times10^{-3}$ & $2.69\times10^{-5}$ & 7.07 & 0.123 \\
AR 13392 & multi, $\lambda_{\rm ff}=10^{-3}$ & $4.22\times10^{-3}$ & $1.79\times10^{-5}$ & 4.21 & 0.073 \\
\hline
\end{tabular}
\end{table}


\bibliography{main}{}
\bibliographystyle{aasjournalv7}



\end{document}